\documentclass[reprint,superscriptaddress,pre]{revtex4-1}
\usepackage{color}
\usepackage{mathtools}
\usepackage{hyperref}
\usepackage{amsmath,amssymb,amsfonts}
\usepackage{graphicx}
\usepackage{dcolumn}
\usepackage{bm}
\usepackage{float}
\usepackage{epstopdf}
\usepackage{textcomp}
\usepackage{verbatim}
\usepackage[normalem]{ulem}

\def\be{\begin{equation}}
	\def\ee{\end{equation}}
\def\bea{\begin{eqnarray}}
	\def\eea{\end{eqnarray}}
\def\bsplit{\begin{split}}
	\def\esplit{\end{split}}

\def\p{\partial} 
\def\nn{\nonumber}
\def\f{\frac}

\def\bs{\boldsymbol}

\def\l{\left(}
\def\r{\right)}

\def\la{\langle}
\def\ra{\rangle}
\def\lla{\left\langle}
\def\rla{\right\rangle}
\def\bs{\boldsymbol}

\def\g{\gamma}

\def\mr{\mathrm}
\def\refn{Eq.\,\ref}

\newcommand{\ncbs}{\affiliation{Simons Centre for the Study of Living Machines, National Centre for Biological Sciences, TIFR, 560065 Bangalore, India}}
\newcommand{\curie}{\affiliation{Laboratoire Physico Chimie Curie, Institut Curie, Université PSL, CNRS UMR168, Sorbonne Université, 75005 Paris, France}}
\newcommand{\pks}{\affiliation{Max Planck Institute for the Physics of Complex Systems, 01187 Dresden, Germany}}

\begin{document}
	
	\graphicspath{{figures/}}

	\title{Stochastic nanoswimmer: a multistate model for enzyme self-propulsion and enhanced diffusion}
	
	\author{Amit Singh Vishen}
	\pks
	\author{Jacques Prost}
    \email{jacques.prost@curie.fr}
	\curie
	\email{jacques.prost@curie.fr}
	\author{Madan Rao}%
	\email{madan@ncbs.res.in}
    \ncbs

	\date{\today}

	\begin{abstract}
		 
Several enzymes exhibit enhanced diffusion in the presence of a substrate. One explanation of this enhancement arises from fluctuating dimer models, which suggest that enzymes have a higher diffusion constant when interacting with substrates compared to when they are free. Another possible mechanism, suggested in both experimental and theoretical studies, is that enzymes act as nanoswimmers. However, existing nanoswimmer models have struggled to account for the exceptionally high self-propulsion speeds observed in experiments, with estimated increases in diffusion due to self-propulsion found to be minimal. In this study, we model enzymes as dimers with fluctuating mobility. We show that even dimers can exhibit run-and-tumble motion when transitioning between states of varying mobility within the enzymatic cycle. By exploring a three-state enzymatic cycle, we identify the conditions under which self-propulsion speeds and increases in diffusion rate consistent with experimental observations can be achieved.
		
	\end{abstract}

	\maketitle
	
	\section{Introduction}
	
	Enzymes are polymers with a complex interplay between form and function. Unlike chemical catalysts, enzymes' catalytic activity is dominated by weak interactions and nanoscale morphological transitions; this leads to enzymes having shape fluctuations at timescales spanning multiple orders of magnitude -- from thermal fluctuations of the small segments at pico to nanoseconds timescale to largescale correlated conformation fluctuations at micro to milliseconds timescale \cite{Henzler-Wildman2007b, Henzler-Wildman2007a, Henzler-Wildman2007, Piephoff2017, Pang2017}.
	In the presence of the substrate, switching between specific conformation states facilitates the catalytic activity of the enzyme converting the substrate into the product. 
	In recent years, the effect of the substrate-induced mechanical deformation of the enzyme on its motion as a whole, has become an active research topic. 
	An increase in diffusivity in the presence of its substrate has been observed experimentally for several enzymes \cite{Dey2014, Dey2015,  Astumian2014, Sengupta2014, Riedel2014, Muddana2010, Jee2020,Jee2018, Wang2020}. The extent of enhancement correlates with the rate of reaction \cite{Jee2020}. Moreover, single-molecule experiments, suggest that enzymes, such as urease and aldolase, undergo run and tumble motion, with run-length of about $100 \,\mr{nm}$, and run time of a microsecond \cite{Jee2018, Slochower2018}. While there is ongoing controversy about the magnitude of the enhanced diffusion observed in several experiments, there is nothing in principle that would disallow an enhancement. Here we explore the conditions under which a diffusion enhancement may occur \cite{Gunther2021comment, Wang2021response}.  
	
	Various promising theoretical models have been proposed to explain enhanced diffusion and self-propulsion. However, a satisfactory mechanistic understanding of the enhanced diffusion of active enzymes is still lacking. 
	In ref.\,\cite{Illien2017,Illien2017b, Adeleke-Larodo2017, Kondrat2019}, the enhancement of diffusion has been explained by postulating that the active enzyme's diffusion coefficient is larger than the inactive enzyme. In the presence of substrate, the enzyme spends more time in the active state; hence this leads to an enhanced diffusion compared to the state without substrate.
	However, it is not clear why an active enzyme has a larger diffusion constant than an inactive enzyme.   
	Self-propulsion of nanoscale enzymes has been proposed \cite{Sakaue2010, Huang2012} applying the general ideas of microswimmers \cite{Najafi2004, Pooley2007,Earl2007,Golestanian2008,Golestanian2008a,Golestanian2009}. Theoretical estimates show that the enhancement in diffusion due to self-propulsion is not large enough to explain the numbers observed in experiments  \cite{Golestanian2015, Zhang2019,Bai2015,Feng2020}. The argument can be summarized as follows. A particle making a stroke of size $\delta l$ at a rate $k$ can swim at most at speed $v \approx k \delta l$. For an enzyme the $\delta l \sim 10^{-9}\, \mr{m}$ and $k\sim 10^4\,\mr{s^{-1}}$ giving $v \sim 10^{-5} \mr{m/s}$. This velocity is two orders of magnitude smaller than the velocities observed in the experiments, $v_\mr{exp}\sim10^{-3} \mr{m/s}$ \cite{Jee2018}. Since the rotational diffusion constant is $D_r \sim 10^5\, \mr{s^{-1}}$, the enhanced diffusion obtained with this estimates would be $10^{-3} \mr{\mu m^2/s}$, which is much smaller than the observed enhancement in diffusion $\sim 10\,\mr{\mu m ^2/s}$.

	In this work, we aim at finding the conditions under which  the above difficulties could be solved.
 We use a simplified model in which the enzyme is modeled as a dimer of two spherical particles interacting through an inter-particle potential. The dimer exists in multiple states with different interaction potentials and different mobilities. 
	The difference in mobility in different states is the first essential addition to the dimer model, which makes it different from the previous theoretical studies of stochastic dimer \cite{Mikhailov2015, Illien2017, Illien2017b}.  
	The second crucial element is the inclusion of multiple enzyme states.  
	
    To emphasize, while it is well understood that a dimer with constant mobility cannot swim, we show here that a dimer, whose mobility and interaction potential are  functions of time, can indeed self-propel. 
	Incorporating multiscale conformation switching of the enzyme, we show that the self-propulsion can enhance diffusion much more than that previously estimated. 
	 We show that the enzyme can perform run-and-tumble motion,  and find the conditions under which consistency with the experiments by Jee et al. \cite{Jee2018} are obtained. Since the enhancement in diffusion relies on self-propulsion, this scheme does not require the active enzyme to have a larger diffusion constant than the enzyme in the passive state.


	\section{Fluctuating dimer}
	
	\begin{figure}
		\centering
		\includegraphics[width= \linewidth]{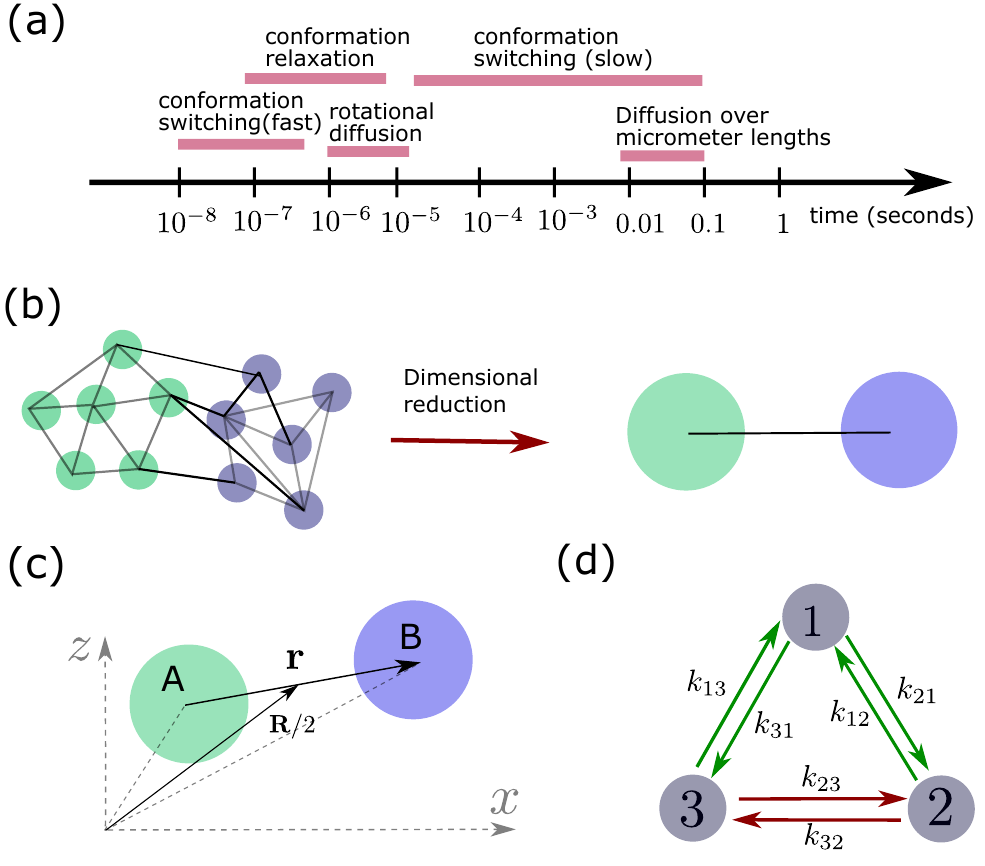}
		\caption{(a) Timeline showing the hierarchy of timescales involved in the enzyme dynamics: from fast conformation fluctuations at nanoseconds to slow fluctuations at milliseconds. (b) Schematic showing the dimensional reduction of a protein with large number of degrees of freedom to a dimer model two degrees of freedom that capture most of the conformation fluctuation. 
    (c) Schematic showing the position $R$ and separation $r$ of the dimer. (d) Schematic showing the three conformation states and the rates of switching between different states.
 }
		\label{fig:schematic}
	\end{figure}
	
	Consider an enzyme $E$ which converts the substrate $S$ into the product $P$. For a fixed concentration of $S$ and $P$, this system is out-of-equilibrium, driven by the imposed chemical potential difference between the substrate and the product. 
	The enzyme fluctuates between different conformational substates (CS). The energy landscape governs the transition between these states with minima separated by a barrier of different heights \cite{Ansari1985, Frauenfelder1991, Lewandowski2015}. 
	The hierarchy in barrier heights also leads to a hierarchy in the timescale of conformation fluctuation (Fig.\,\ref{fig:schematic}(d)), which varies from picoseconds to seconds \cite{Henzler-Wildman2007b, Henzler-Wildman2007a, Henzler-Wildman2007}.
	Its interaction with the substrate biases the rate of switching between different conformations. 
	
	During the enzymatic cycle the enzyme switches from one state two another which can be represented as a motion along a reaction coordinate. This reaction coordinate can be mapped to the real space motion of different segments of the enzyme. 
 We assume here that the motion of the enzyme in the real spaces can be captured with few slow degrees of freedom.
	
	In this work, we use a simplified model of protein kinetics.
	Following earlier works \cite{Mikhailov2015,Adeleke-Larodo2017} we model the enzyme $E$ as a dimer of spherical colloids with centers at position ${\bf R}_\mr{A}$ and ${\bf R}_\mr{B}$, respectively (see Fig.\,\ref{fig:schematic} (a)), interacting through an interparticle potential. The enzyme can exist in different conformational states (CS) labeled by subscript $\gamma$ \cite{English2006, Lu2014,Piephoff2017}.
	Different CS have different interaction energy $\mr{U_{\gamma}}(r)$, and different mobility of the colloids. For simplicity we take one of the colloids of fixed mobility $\mu_\mr{A}$ and other with state-dependent mobility $\mu_{\gamma}$. 
	A two-state dimer with different interaction energy in different states has been analyzed before. The crucial new element in our work is the introduction of state-dependent mobility. In different CS, the interaction of the enzyme with water can be very different. Differences in the arrangement of hydrophilic and hydrophobic amino acids may lead to differences in the boundary condition or hydration shell, resulting in different mobilities in different states \cite{Zhang2007,Dahanayake2018}.

	The stochastic dynamics of the colloids for the above model, in the overdamped limit, is given by the following Langevin equation:
	\bea
	\label{eq:langevin_a}
	\f{d {\bf R}_\mr{A}}{dt}&=& -\mu_\mr{A} \nabla_{{\bf R}_\mr{A}} \mr{U_{\gamma}}(r) + \sqrt{2T\mu_\mr{A}}\,{\bs \xi}_\mr{A},\\
	\label{eq:langevin_b}
	\f{d {\bf R}_\mr{B}}{dt} &=& -\mu_{\gamma} \nabla_{{\bf R}_\mr{B}} \mr{U_{\gamma}}(r) + \sqrt{2T\mu_{\gamma}}\,{\bs \xi_\mr{B}},
	\eea 
	where $T \equiv k_B T$ is temperature times the Boltzmann constant, ${\bs \xi_\mr{A}}$ and ${\bs \xi}_\mr{B}$ are Gaussian white noise vectors with correlation $\langle \xi_{\mr{A} i}(t)\xi_{\mr{B} j}(t')\rangle = \delta_\mr{AB}\delta_{ij}\delta(t-t')$, with $i$ and $j$ labeling the spatial coordinates, $\mu_\mr{A}$ and $\mu_{\gamma}$ are the mobilities, and $\mr{U_{\gamma}}$ is the interaction potential between the two colloids in the state $\gamma$, where $\gamma \in \{1,2,...N\}$ and $N$ is the number of CS. 
	
	Writing \refn{eq:langevin_a} and \refn{eq:langevin_b} in terms of the sum of the position vectors ${\bf R} \equiv {\bf R}_\mr{A} + {\bf R}_\mr{B}$, and separation coordinate ${\bf r}\equiv {\bf R}_\mr{B} - {\bf R}_\mr{A}$,  we get
	\bea 
	\label{eq:langevin_R}
	\f{d {\bf R}}{dt}&=& -(\mu_{\gamma} -\mu_\mr{A}) \f{ \p \mr{U_{\gamma}}}{\p r}\,\hat{\bf n}+ \sqrt{2T\mu_\mr{A}}{\bs \xi_{A}} + \sqrt{2T\mu_{\gamma}}\,{\bs \xi_B},\quad \\
	\label{eq:langevin_r}
	\f{d {\bf r}}{dt} &=& -(\mu_{\gamma} + \mu_\mr{A})\f{ \p \mr{U_{\gamma}}}{\p r}\,\hat{\bf n} + \sqrt{2T\mu_{\gamma}}{\bs \xi_B} - \sqrt{2T\mu_A}\,{\bs \xi_A},
	\eea 
	where $\hat {\bf n}$ is the unit vector along ${\bf r}$.
	The probability density function (PDF) in the state $\gamma$ obtained using the Langevin dynamics \refn{eq:langevin_R} and \refn{eq:langevin_r} and including the switching between states has the form \cite{Gardiner,Julicher1997}
	\be
	\label{eq:prob}
	\p_t \mr{P}_{\gamma}({\bf R,r},t) = -\nabla \cdot {\bf J}_{\gamma} + \sum_{\delta = 1}^{N}k_{\gamma\delta} \mr{P}_{\delta} -  k_{\delta\gamma} \mr{P}_{\gamma},
	\ee 
	where $k_{\delta\gamma}$ is the transition rate from state $\gamma$ to state $\delta$. 
    The divergence of the probability current is given by 
	\bea 
	\label{eq:flux1}
	\nn -\nabla \cdot {\bf J}_{\gamma}  &=& \nabla_{{\bf R}} \cdot\l  \Delta \mu_{\gamma} \f{\p\mr{U_{\gamma}}}{\p r}\,\hat{\bf n} + T   \Delta \mu_{\gamma} \nabla_{{\bf r}}   + T \bar \mu_{\gamma} \nabla_{{\bf R}} \r \mr{P}_{\gamma} \\
	&+& \nabla_{{\bf r}} \cdot \l   \bar \mu_{\gamma} \f{\p \mr{U_\gamma}}{\p r}\,\hat{\bf n}  + T \Delta \mu_{\gamma}  \nabla_{{\bf R}}  +  T \bar \mu_{\gamma} \nabla_{{\bf r}}\r \mr{P}_{\gamma},\quad \,\,
	\eea 
	where we have defined $\Delta \mu_{\gamma} \equiv \mu_{\gamma} - \mu_\mr{A}$ and $\bar \mu_{\gamma} \equiv \mu_{\gamma} +\mu_\mr{A}$.
    In Appendix \ref{appendix-dimer} we present \refn{eq:prob} in Cartesian and spherical coordinates.
 
	If the detailed balance relation is satisfied, i.e., $k_{\gamma\delta}/k_{\delta\gamma}= e^{-\l\mr{U_{\gamma}}-\mr{U_{\delta}}\r/T}$ \cite{Gardiner}, \refn{eq:prob} leads to an equilibrium steady-state, i.e., $\mr{P}_{\gamma} \propto e^{-\mr{U}_{\gamma}/T}$. 
	As mentioned above, the enzyme kinetics is driven out of equilibrium due to the steady flux of substrate  into the system and flux of product out of the system.

	In this work we analyze a three-state model. State $1$ corresponds to a passive state of the enzyme (unbound state). States $2$ and $3$ are active states that the enzymes reach on interacting with the substrate (bound states).  A cycle through states $1$, $2$, and $3$ converts one substrate into a product (see Fig.\,\ref{fig:schematic} (b). 
	
	Given the complexity of the enzymatic process, relating the non-equilibrium drive from the fixed chemical potential difference of substrate and product to the rates of switching between different states is far from obvious \cite{Hammes2009, Csermely2010}. Although, in general, the transition rates can be a function of ${\bf R}$ and ${\bf r}$; here, for simplicity, we take the switching rates between different states to be a constant. This necessarily implies that all the transitions are out-of-equilibrium unless $\mr{U}_{\gamma}$ is a constant.
	\refn{eq:prob}, written explicitly for the three states reads 
	\bea
	\label{eq:prob1}
	\p_t \mr{P}_1 &=& -\nabla \cdot {\bf J_1} + k_{12} \mr{P}_2 + k_{13} \mr{P}_3 -  (k_{21} + k_{31}) \mr{P}_1, \\ 
	\label{eq:prob2}
	\p_t \mr{P}_2 &=& -\nabla \cdot {\bf J_2} +  k_{21} \mr{P}_1 + k_{23} \mr{P}_3 - (k_{12}+ k_{32}) \mr{P}_2 , \\
	\label{eq:prob3}
	\p_t \mr{P}_3 &=& -\nabla \cdot {\bf J_3} +  k_{32} \mr{P}_2 + k_{31} \mr{P}_1 - (k_{23}+ k_{13}) \mr{P}_3,
	\eea 
	where $\mr{P}_1$ is the PDF of the unbound state, $\mr{P}_2$ and $\mr{P}_3$ are the PDF's of the intermediate bound states.

	There are four timescales in the problem: (i) transition rates between the bound and unbound state: $k_{b\leftrightarrow u} \in \{k_{12},k_{21},k_{13},k_{31}\}$, (ii) transition rates in-between the bound states: $k_{b\leftrightarrow b} \in \{k_{23},k_{32}\}$, (iii) rotational diffusion of the dimer: $D_\mr{rot}$, and (iv) the relaxation rate of the separation variable $r$: $k_\mr{sep} = \bar \mu_{\gamma} k_{\gamma}$.

	The timescale of switching between bound and unbound states is obtained from the enzymes' reaction velocity studied in Ref.\,\cite{Jee2020}; this is on the order of $k_{b\leftrightarrow u} \sim 1-10^{4}\,\mr{s}^{-1}$. 
	The fluctuation of the protein between CS is hierarchical, spanning timescales from milliseconds to nanoseconds. We are interested in the fluctuations at intermediate timescales of microseconds; hence, we consider $k_{b \leftrightarrow b} \sim 10^{6}\,\mr{s}^{-1}$. Thus we introduce a separation of timescales between the dynamics of binding and unbinding and switching between the bound states, i.e., $k_{b \leftrightarrow b} \gg k_{b\leftrightarrow u}$. 
	
	To estimate the relaxation timescale of $r$, we consider a harmonic interparticle interaction $\mr{U}_{\gamma}(r) = k_{\gamma}(r-l_{\gamma})^2$, with a 
	stiffness $k_{\gamma}  \sim 10^{-(3-4)} N/m$ \cite{Lewalle2008,Alemany2016, Sonar2020,Heidarsson2021}, and $l_{\gamma}\sim 10\,\mr{nm}$. The mobility of ``monomer" $\mr{A}$ is $
	\mu_\mr{A} \sim 1/6\pi \eta a$; thus for a radius $a$  $\approx 1-10\,\mr{nm}$, we get $ \mu_\mr{A} \sim 10^{9-10}/ \mr{Pa\,m\,s}$.
 Note that the mobility could be much smaller due to contribution from the internal friction. This, for instance, is the case for molecular motors where changing the viscosity of the fluid by a factor hundred does not change the motor velocity, when the motor density is high enough \cite{Hunt1994}. Taking $\mu_{\gamma}\sim \mu_\mr{A}$, the timescale of relaxation of $r$ is $ k_\mr{sep} = \mu_{\gamma} k_{\gamma} \sim 10^{5-7}\, \mr{s}^{-1}$. 
	The translation diffusion constant is $D_t = T\mu_1\mu_{\gamma}/(\mu_1+\mu_{\gamma})  \sim 50\,\mu m^2/s$, and the rotational diffusion constant is on the order of $D_\mr{rot} \sim T(\mu_1+\mu_{\gamma})/l_{\gamma}^2 \sim 10^{5-6}\,\mr{s}^{-1}$. 
	Thus we see that there is a clear hierarchy of timescales between the four rates: $ k_\mr{sep},k_{b\leftrightarrow b} \gg D_\mr{rot} \gg k_{b\leftrightarrow u}$.
	
 For a spatially constant density profile of substrate and product, the dynamics are independent of the enzyme position ${\bf R}$. However, when a spatial gradient of the substrate density is imposed, the rates $k_{21}$ and $k_{31}$ that depend on the substrate and product concentration, respectively, will now depend on ${\bf R}$. For a spatial gradient over length scale $l_s$,  the rates can be considered effectively constant over a timescale less than $1/\tau_t = l_s^2/D_t$. In the following, we take the substrate gradient to vary at length scales larger than a few microns; hence, we take the rates to be constant.

	\section{Dimer as a self-propelled particle}
	
	In the following, we obtain the effective dynamics of the dimer by integrating out the fast degrees of freedom---separation of the dimer and the intermediate bound states---to get effective dimer dynamics with a reduced set of variables---dimer switching between the bound and unbound state, position ${\bf R}$, and orientation.

	The timescale difference between $k_{b\leftrightarrow b}$ and $k_{b\leftrightarrow u}$ allows us to obtain an effective two-state model from the three state model. 
	To integrate out the fast intermediate bound states, we define $\mr{P_b}= \mr{P}_2+\mr{P}_3$ and $\mr{P_d}= \mr{P}_2-\mr{P}_3$. The dynamics of $\mr{P_b}$ and $\mr{P_d}$ are obtained by adding and subtracting \refn{eq:prob2} and \refn{eq:prob3}. 
	It can be seen from the equations that $k_{b\leftrightarrow b} \gg k_{b\leftrightarrow u}$ implies fast relaxation of  $\mr{P_d}$.  Using $k_{23},k_{32}\gg k_{12},k_{21}$, at steady state we get $\mr{P_d} \simeq (k_{23}-k_{32})\mr{P_{b}}/(k_{32}+k_{23})$. Substituting this in the equation for $\mr{P_b}$ we get
	\bea 
	\label{eq:probs}
	\p_t \mr{P}_b &=& -\nabla \cdot {\bf J_b} +  k_{bu} \mr{P}_u  - k_{ub} \mr{P}_b, 
	\eea 
	where ${\bf J_b}=\l {\bf J_2}+{\bf J_3} \r$ and the effective binding and unbinding rates are 
	$k_{ub} = (k_{23}k_{12}+k_{32}k_{13})/(k_{23} + k_{32})$ and $k_{bu}=(k_{21} + k_{31})$. 
	We now have an effective two-state dynamics, where state $u$ represents the unbound state (state one is relabeled to state $u$), and state $b$ represents the bound state. 
	After integrating \refn{eq:prob1} and \refn{eq:probs} over $r$ and the intermediate bound states two and three, the effective dynamics of orientation ${\bf \hat n}$ and  position ${\bf R}$ reads
	\bea 
	\label{eq:eff_prob}
	\nn \p_t \tilde P_{\gamma}  &=& \nabla_{\bf R} \cdot \l {v}_{\gamma}\, {\bf \hat n} +  T\delta\mu_{\gamma}{\bf\hat n} \l  {\bf\hat n} \cdot \nabla_{\bf R}\r\r \tilde P_{\gamma} \\
	\nn &+&  \f{\p}{\p\theta} \l \lla\f{T\Delta \mu}{l}\rla_{\gamma}{\bf \hat n}\times \nabla_{\bf R} + \lla \f{T\bar \mu}{l^2} \rla_{\gamma} \f{\p}{\p\theta}\r \tilde P_{\gamma} \\
	&+& T\bar \mu_{\gamma} \nabla_{\bf R}^2 \tilde P_{\gamma} + k_{\gamma\delta} \tilde P_{\delta}- k_{\delta \gamma} \tilde P_{\gamma},
	\eea 	
	where $ \tilde P_{\gamma}({\bf R, \hat n}) \equiv \int_{0}^{\infty} dr\,r^d\,\mr{P}_{\gamma}({\bf R, \hat n},r) $, $d$ is the dimensionality, for $\gamma \in \{ u,b\}$, $v_{\gamma}$ and $\delta \mu_{\gamma}$ are defined by the relation: $ v_{\gamma}\tilde P_{\g}+\delta \mu_{\gamma}\,{\bf \hat n} \cdot \nabla_{\bf R} \tilde P_\g = \int_{0}^{\infty} dr\,r^d\, \l \p_r \mr{U} - T/r\r_{\gamma}$. The averages are defined as $\la \phi \ra_u = \phi_1$ and $\la \phi \ra_b = \l k_{23}\phi_2 + k_{32}\phi_3\r/(k_{23}+k_{32}) $. See Appendix \ref{appendix-effectivedynamics} for the derivation of \refn{eq:eff_prob}. 

For small deformations we can take the interaction potential to be harmonic, i.e.,  $\mr{U_{\gamma}} = k_{\gamma}(r-l_{\gamma})^2/2 $. Furthermore, to leading order, we get $1/r \approx  1/l_\gamma + (r - l_\gamma)/l_\gamma^2 $. With these assumptions we only need to solve for the average of  $r$ to evaluate the integral $\int_{0}^{\infty} dr\,r^d\, \l \p_r \mr{U} - T/r\r_{\gamma}$. 
To obtain the dynamics of the average separation $\la r \ra$, we multiply
 \refn{eq:prob1}--\refn{eq:prob3} by $r$ and integrate. The resulting dynamics reads:
	\bea
	\label{eq:sep1}
	\p_t\la r_1\ra &=& - \bar \mu_1  \tilde k_1(\la r_1 \ra -d_1 \tilde l_1) - T\Delta\mu_1\, {\bf \hat n} \cdot \nabla_{\bf R} \tilde P_1, \\ 
	\label{eq:sep2}
	\nn\p_t \la r_2 \ra &=& - \bar \mu_2  \tilde k_2(\la r_2 \ra -d_2 \tilde l_2) - T\Delta\mu_2\, {\bf \hat n} \cdot \nabla_{\bf R}  \tilde P_2 \\
	&+& k_{23} \la r_3\ra -  k_{32} \la r_2\ra, \quad \\
	\label{eq:sep3}
	\nn\p_t \la r_3 \ra  &=&-\bar \mu_3  \tilde k_3(\la r_3 \ra -d_3 \tilde l_3) - T\Delta\mu_3\, {\bf \hat n} \cdot \nabla_{\bf R} \tilde P_3 \\
	&+& k_{32} \la r_2 \ra - k_{23} \la r_3\ra,
	\eea   
	where the renormalized stiffness and length of the spring are $\tilde k_{\gamma} = \l k_{\gamma} + T/l_{\gamma}^2\r$ and $\tilde l_{\gamma} = \l1+T/(k_{\gamma}l_{\gamma}^2+T)\r$.
	The correction term $T/l_{\gamma}^2 \sim 10^{-5} N/m$,  which is one to two orders of magnitude smaller than $k_{\gamma}$. Thus we can take $\tilde k \approx k$ and $\tilde l \approx l$. 
 From the definition $k(\la r\ra_\gamma - l_{\gamma}) = v_{\gamma}\tilde P_{\g}+\delta \mu_{\gamma}\,{\bf \hat n} \cdot \nabla_{\bf R} \tilde P_\g$, from \refn{eq:sep1}
	at steady state, we get
	$v_1 = 0$ and $\delta \mu_1 = \Delta \mu^2_1/\bar \mu_1$. And from \refn{eq:sep2} and \refn{eq:sep3}, we get
	\be
	\label{eq:vel}
	v_b = \f{2k_{23}k_{32}\tilde k_2\tilde k_3\mu_a(\bar\mu_2-\bar\mu_3)\delta l}{(k_{23}+k_{32})\l \bar\mu_2 \tilde k_2\bar\mu_3 \tilde k_3 + k_{23}\bar\mu_2 \tilde k_2  + k_{32} \bar \mu_3 \tilde k_3 \r }.
	\ee
	Thus we see that for $\delta l \equiv l_2 - l_3 \neq 0$ and $\bar\mu_2 \neq \bar\mu_3$, in the bound state, the dimer has a finite velocity along the orientation vector.
	The sign of the velocity is set by the product $(\mu_2 - \mu_3)(l_2-l_3)$, it is positive if the state with the larger mobility is the state with a larger length. 
 	
	For $ \bar \mu r_2,\bar \mu r_3\gg k_{32},k_{23}$,  $\mu_2\gg \mu_a \gg \mu_3$, and $k_{23} = k_{32} = 1/\tau$  we get 	$ v_b \approx \delta l/\tau$. This shows that the enzyme can self-propel in the bound state with a stroke rate set by the intermediate switching rate. This is what one gets from the order-of-magnitude estimates \cite{Golestanian2008}, except here the timescale is the timescale of the switching between intermediate bound states which is taken to be much larger than switching between bound and unbound states. For $\tau^{-1}\sim 10^{6}\, \mr{s^{-1}}$ and $\delta l \sim 1\, \mr{nm}$, we get $v_s \sim 10^{-3}\, \mr{m/s}$. This is of the same order of magnitude as reported in Jee et. al. \cite{Jee2018}.

	\begin{figure}
		\centering
		\includegraphics[width= 0.7\linewidth]{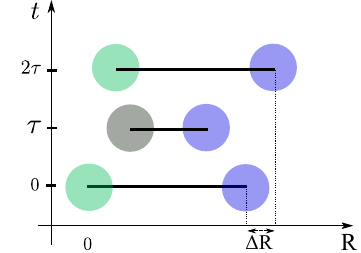}
		\caption{ Schematic showing the translation of the dimer on cycle deformation with switching mobility. The mobility and the interaction potential between the spherical particles switches at after a fixed time interval $\tau$. In one cycle the center of the dimer moves a distance $\Delta R$. The velocity $v = \Delta R/2\tau$ is given by \refn{eq:det_v}.}
		\label{fig:det_cycle}
	\end{figure}
	
	\subsection{Toy model of self-propulsion: deterministic dynamics}
	
	We can understand clearly this self-propulsion mechanism by looking at the one-dimensional deterministic dynamics of the dimer. 
	Consider two spheres positioned at $R_A$ and $R_B$, respectively, interacting by a potential that switches periodically between the two values $ \mr{U}_3=k_3(r-l_3)^2/2$ and $\mr{U}_2=k_2(r-l_2)^2/2$, where $r= R_B - R_A$, at a time interval $\tau$. Concomittantly, let the mobility of the colloid at $R_B$ switch between $\mu_3$ and $\mu_2$, whereas the mobility of colloid at $R_A$ is fixed at a constant value $\mu_A$. The deterministic dynamics of the two colloids in the overdamped limit, as given by \refn{eq:langevin_R} and \refn{eq:langevin_r}, reads
	\be
	\label{eq:det_dyn}
	\f{d R}{dt}= (\mu_A -\mu_{\gamma}) k_{\gamma}(r - l_{\gamma}),\,\,\mr{and} \,\, 
	\f{d r}{dt} = -(\mu_A + \mu_{\gamma}) k_{\gamma}(r - l_{\gamma}).
	\ee 
	The displacement of the dimer position $R$ obtained from \refn{eq:det_dyn} reads
	\be 
	\label{eq:pos_dimer}
	R(t) - R(0) = \f{\mu_A - \mu_{\gamma}}{\mu_A + \mu_{\gamma}}\l r(t) - r(0)\r. 
	\ee 
	At time $t=0$ the potential changes from  $\mr{U}_3$ to $\mr{U}_2$, and the mobility changes from $\mu_3$ to $\mu_2$. In the limit $\mu_a k_{\gamma},\mu_{\gamma} k_{\gamma}\gg \tau$, the dimer relaxes to the new length $l_2$. 
	From \refn{eq:pos_dimer} we get that the dimer  moves by distance $\Delta R_3 = (l_2-l_3)(\mu_A -\mu_2)/(\mu_A + \mu_2)$ .
	At the time $t = \tau$ the potential and the mobility switch back to  $\mr{U}_3$ and $\mu_3$, respectively. The dimer now relaxes to the new equilibrium length $l_3$, and the dimer moves a distance $\Delta R_2 = (l_3-l_2)(\mu_A - \mu_3)/(\mu_A + \mu_3)$. The total displacement in one cycle of time $2\tau$ is $\Delta R = \Delta R_3 + \Delta R_2$, and the average velocity $v_0 = \Delta R/2\tau$ reads
	\be 
	\label{eq:det_v}
	v_0 = \f{1}{\tau}\l \f{\mu_A(\mu_2-\mu_3)}{(\mu_2+\mu_A)(\mu_3+\mu_A)}\r(l_2 - l_3) .
	\ee 
	This expression is identical to the velocity in \refn{eq:vel} in the limit of slow switching ($k_2 \bar \mu_2 , k_3\bar \mu_3 \gg k_{23}, k_{32}$)  and $k_{23} = k_{32}$. 
	
	We have studied the effects of hydrodynamic interaction between the two particles on the effective velocity of the dimer in Appendix \ref{appendix-DetDyn}. We get an expression for velocity which is larger in value than \refn{eq:det_v}, but does not change the order of magnitude.
	
	\section{Effective diffusion constant}

	To compute the effective diffusion of the enzyme at long times, we need to integrate out the orientation ${\bf \hat n}$ and the transitions between bound-unbound states from \refn{eq:eff_prob}. 
	A reasonable limit for the enzyme is that the rotational diffusion is faster than the switching time. 
	In this limit, one gets a particle with an effective diffusion constant in both the bound and unbound states.  
	Treating the orientation vector as a fast variable, we can adiabatically eliminate the fast modes. The moments of the orientation are $p_{mn}^\g = \int_{0}^{2\pi} d\theta \cos^m\theta\sin^n\theta \tilde P_{\gamma}$. The zeroth moment is the density: $\rho_\gamma = p_{00}^\g = \int_{0}^{2\pi} d\theta \,\tilde P_{\gamma}$.
	As shown in Appendix \ref{appendix-DiffConst}  to solve for the density it suffices to consider only the first and second moments. 
	The effective density dynamics in the unbound state obtained by integrating out the orientation from \refn{eq:eff_prob} reads  
	\be
	\p_t \rho_u = D_u  \nabla^2 \rho_u - k_{bu} \rho_u + k_{ub}\rho_b,
	\ee
	where $D_u$ is the effective diffusion constant. As expected for the dynamics of a dimer at equilibrium, $D_u = 4T\mu_a \mu_1/(\mu_a + \mu_1)$. 
 Similarly, the effective density dynamics in the bound state reads
	\be 
	\p_t \rho_b = \l \f{v_b^2}{D_\mr{rot}}+ D_b \r \nabla^2 \rho_b  + k_{bu} \rho_u - k_{ub}\rho_b,
	\ee 
	where
	\be 
	D_{b} = \f{T}{2}\l 2\la \bar \mu \ra  - \f{1}{(\la \bar\mu/l^2\ra)}\lla \f{\Delta\mu}{l}\rla^2 - \delta\mu_b \r,
	\ee 
	is the diffusivity of the enzyme in the bound state without self-propulsion. Its value depends on the rate of switching between the intermediate states (see Appendix \ref{appendix-DiffConst}). 
	
	Finally, the effective diffusivity at long times obtained by integrating out the switching dynamics is simply 
	\be 
	\label{eq:densityE}
	\p_t \rho =  \nabla^2 \l D_\mr{eff} \rho \r,
	\ee 
	where
	\be 
	D_\mr{eff} = \f{k_{ub} D_u + k_{bu} D_b}{k_{bu} + k_{ub}} + \f{ k_{bu} v_b^2/D_\mr{rot}}{k_{bu} + k_{ub}} .
	\ee 
	The first term has been proposed as a diffusion enhancement mechanism \cite{Illien2017,Illien2017b, Adeleke-Larodo2017}; it requires that the diffusivity in the bound state is larger than that of the unbound state. The second term is due to the directed motion in the bound state. 
	The diffusion is enhanced if both terms are positive. If the diffusion in the bound state is lower than in the unbound state, then the sign of the first term is negative, and the effective diffusion may decrease for a large enough difference. 
	
	For $\tilde \mu_s \approx \tilde \mu_1$, the effective diffusion constant is $D_\mr{eff}/D_u -1 =  v_b^2/D_\mr{rot}D_u \approx 10 \%$, which is on the order of the maximum observed enhancement in diffusion \cite{Jee2020}. 
	The effective drift term is given by
	${\bf v}_\mr{eff} = \nabla D_\mr{eff}$.
	The steadystate distribution corresponding to  \refn{eq:densityE} is given by 
	$\rho_{ss} \propto 1/D_\mr{eff}$.
	The unbinding rate $k_{ub}$ can be taken to be constant; however, the binding rate $k_{bu}$ depends on the concentration of the substrate, which can be space dependent.
	Taking the linear dependence of the binding rate of the form $k_{bu} = k_0 S$, where $S$ is the substrate concentration.  We see that the effective diffusion is a monotonically increasing function of $S$. Consequently, the PDF of the particle position is a monotonically decreasing function of $S$. 
	We get an anti-chemotactic response. 
    In previous works \cite{Jee2018, Jee2018b}, this behavior has been 
 attributed to the position-dependent diffusion constant; we get the same result here. Finally, we estimate the energy dissipated into the fluid by the swimming enzyme. The velocity field created by the enzyme swimming goes as $v_0 l^3/r^3$. The dissipation in the fluid during the self-propulsion phase is given by
	\be 
	\dot E_f \sim \int_V \eta (\nabla v)^2 dV \sim \eta v_0^2 \int_l^{\infty} \f{l^6}{r^6} dr \sim \eta v_0^2 l .
	\ee 
	For typical numbers, 
 $\eta \approx 10^{-3}Pa.s$, $v_0 \approx 1 mm/s$, and $l \approx 10 nm $ this gives $\dot E_f \approx 10^{-17} J/s$, for a self-propulsion state of about a microsecond we get $\Delta E_f \approx k_BT$, which is in the range of $\Delta G$ of the enzymes showing enhanced diffusion \cite{Jee2020}. 
	
	\section{Discussion}
	
	To summarize, we present a novel model of stochastic enzyme swimming, which relies on the changes in the frictional coupling between the particle and the medium, due to out-of-equilibrium enzymatic activity. Different friction in different states of the enzyme may be due to changes in its geometry and/or changes in the local boundary condition. The latter may result from the exposure of a hydrophobic region of a molecule during a conformational change of a molecule or molecular aggregate.  
	It is well established that in the low Reynolds number limit two rigid particles interacting through an inter-particle potential cannot swim. This is due to the Scallop theorem \cite{Purcell1977}, which forbids the self-propulsion of an object with just one mechanical degree of freedom. 
	A minimal model with beads and springs is the Najafi-Golestanian swimmer with three beads and two springs \cite{Najafi2004,Golestanian2008,Golestanian2008a}. 
 Here we find that with fluctuating mobility, even two beads connected by spring can swim, given the appropriate dynamics of the spring and mobility. This analysis holds resemblance with the active elastic dimer analyzed in Ref. \cite{Kumar2008} that studies the dynamics of a dimer on a substrate, with the monomers subject to a nonequilibrium noise, and with a friction coefficient on one particle depending on the stretch, mimicking the
state dependent mobility. It relates three quantities – mean drift velocity, correlation of internal coordinate and drift velocity, and mean internal force -- each of which can be nonzero only away from thermal equilibrium.
	
	We analyze, in detail, the behavior of a dimer which can exist in three conformations grouped into two subgroups: one unbound state and a bound state with two intermediates. 
	The dimer models an enzyme which is undergoing cycles through the three states in the presence of the substrate. The out-of-equilibrium aspect is obtained by choosing transition rates independent of geometry which automatically breaks detailed balance. We find that the enzyme performs a run-and-tumble motion, with self-propulsion in the bound state and free diffusion in the unbound one.
	For parameters in the physiological range, we show that the enzyme can self-propel with a high speed of the order or millimeter per second, consistent with the experimentally measured velocities \cite{Jee2018}. 
	The timescale relevant for the self-propulsion is the fast switching between the intermediate conformations. However, for these findings to be compatible with the experimental data, such processes have to be repeated a large number of times during an enzymatic cycle. The enzyme has to perform multiple swimming strokes, typically on the order of a hundred in one enzymatic cycle. Under such conditions the enzyme could move multiple times its size in one enzymatic cycle, which corresponds to the experimental findings of Ref. \cite{Jee2018, Slochower2018}. While there is ongoing controversy about the magnitude of the enhanced diffusion and self-propulsion of an enzyme \cite{Gunther2021comment, Wang2021response}, there is nothing in principle that disallows such an enhancement. Here, we have investigated the conditions under which an enzyme can show diffusion enhancement due to substrate mediated self-propulsion. 

	While the simplified model of the enzyme as a dimer is instructive in understanding the enhanced diffusion and self-propulsion, it does not include many other effects that are important for quantitative comparison. Our model can be easily extended to include hydrodynamic interactions \cite{Adeleke-Larodo2017,Illien2017,Vishen2019}, opening a new direction on 
  the collective hydrodynamic effect of enzymes \cite{Hosaka2020,Hosaka2020a,Cressman2008}.
	In different experiments, both chemotactic and anti-chemotactic response has been observed. 	
	This has been reconciled by including diffusophoretic effects  \cite{Mohajerani2018,Agudo-Canalejo2018,Banigan2016}. These effects can be layered on top of the analysis presented here. 
	A more detailed modeling, including the interaction between amino acids and water, needs to be done to account for the mobility fluctuation. Such analysis would be a promising future direction.

	\section{Acknowledgments}
	
	We thank M. Wyart, S. Granick, R. Golestanian, and P. Sens for useful discussions. 
	ASV acknowledges support from the grants ANR-11-LABX-0038, ANR-10-IDEX-0001-02 during his stay at Institute Curie.
	MR acknowledges support from the Department of Atomic Energy (India), under project no.\,RTI4006, the Simons Foundation (Grant No.\,287975), and a JC Bose Fellowship from DST-SERB (India). 
	
	\bibliographystyle{apsrev4-1}
	\bibliography{StochasticSwimmer}{}

	\newpage
	\widetext
	\appendix

	\section{Stochastic dynamics of the dimer in two dimensions}\label{appendix-dimer}

	The Fokker-Planck equation in covariant form, for a diffusion matrix constant in time, reads \cite{Risken}
	\be 
	\label{eq:covarient_FP}
	\f{\p}{\p t} P = \sqrt{|D|}\f{\p}{\p x_i} \f{1}{\sqrt{|D|}}\l -\bar F^i P + D^{ij}\f{\p}{\p x_j} P\r,
	\ee 
	where $|D|$ is the determinant of the diffusion matrix ${\sf D}$, $P$ is a scalar related to the probability density function $\bar P$ by the relation $P = \sqrt{Det}\bar P$, and  
	\be 
	\bar F_i =  F^i+ \sqrt{|D|} \f{\p }{\p x^j} \f{D^{ij}}{\sqrt{|D|}},
	\ee 
	where $F_i$ is the drift term of the corresponding Langevin equation. 
	The Fokker-Planck equation, in Cartesian coordinates, corresponding to the Langevin dynamics given by Eq. (3) and Eq. (4) in the main text reads
	\bea 
	\label{ap:fp}
	\nn \p_t P &=& \f{\p}{\p R_x} \l \Delta \mu \f{ \p\mr{U_{\gamma}}}{\p r}\cos\theta + T\Delta\mu \f{\p}{\p r_x} + T\bar \mu\f{\p}{\p R_x} \r P + \f{\p}{\p R_y} \l \Delta \mu \f{ \p\mr{U_{\gamma}}}{\p r}\sin\theta + T\Delta\mu \f{\p}{\p r_y} + T\bar \mu\f{\p}{\p R_y} \r P \\
	&+&\f{\p}{\p r_x} \l \bar \mu \f{ \p\mr{U_{\gamma}}}{\p r}\cos\theta + T\Delta\mu \f{\p}{\p R_x} + T\bar \mu\f{\p}{\p r_x} \r P + \f{\p}{\p r_y} \l \bar \mu \f{ \p\mr{U_{\gamma}}}{\p r}\sin\theta + T\Delta\mu \f{\p}{\p R_y} + T\bar \mu\f{\p}{\p r_y}\r P,
	\eea 
	where $\Delta\mu = \mu_b - \mu_a$ and $\bar\mu = \mu_a + \mu_b$.
	We now transform \refn{ap:fp} from Cartesian coordinates $r_x$ and $r_y$ to polar coordinates $r$ and $\theta$ using \refn{eq:covarient_FP}. 
	The drift vector and the diffusion matrix are given by
	\be 
	\label{ap:diff_mat}
	{\sf F}_\mr{polar}  = 
	-\f{\p \mr{U}}{\p r}\begin{bmatrix}
		\Delta \mu \cos\theta \\ \Delta \mu\sin\theta \\ 0 \\ \bar \mu \\
	\end{bmatrix}, \quad	{\sf D}_\mr{polar}  = T\begin{bmatrix}
		\bar\mu & 0   & -\f{\Delta\mu}{r}\sin\theta &  \Delta\mu\cos\theta\\
		0 & \bar\mu  & \f{\Delta\mu}{r}\cos\theta & \Delta\mu\sin\theta \\
		-\f{\Delta\mu}{r}\sin\theta& \f{\Delta\mu}{r}\cos\theta &  \f{\bar\mu}{r^2} & 0 \\
		\Delta\mu\cos\theta & \Delta\mu\sin\theta & 0  & \bar\mu
	\end{bmatrix}.
	\ee
	Substituting \refn{ap:diff_mat} into \refn{eq:covarient_FP} we get the following Fokker-Planck equation in polar coordinate: 
	
	\bea 
	\label{ap:fp_polar1}
	\nn \p_t P_{\gamma} &=& \f{\p}{\p R_x} \l \Delta\mu_{\gamma} \l \f{ \p\mr{U_{\gamma}}}{\p r} \r\cos\theta + T\Delta\mu_{\gamma}  \cos\theta \f{\p}{\p r} - T\Delta\mu_{\gamma} \f{\sin\theta}{r}\f{\p}{\p \theta} + T\bar\mu_{\gamma}\f{\p}{\p R_x} \r P_{\gamma} \\
	\nn &+& \f{\p}{\p R_y} \l \Delta\mu_{\gamma} \l \f{ \p\mr{U_{\gamma}}}{\p r} \r\sin\theta + T\Delta\mu_{\gamma}  \sin\theta \f{\p}{\p r} + T\Delta\mu_{\gamma} \f{\cos\theta}{r}\f{\p}{\p \theta}  + T\bar\mu_{\gamma}\f{\p}{\p R_y} \r P_{\gamma} \\
	\nn &+& \f{T\Delta\mu_{\gamma}}{r}\f{\p }{\p \theta}\l -\sin\theta \f{\p P_{\gamma}}{\p R_x} + \cos\theta \f{\p P_{\gamma}}{\p R_y}\r + \f{T\bar\mu_{\gamma}}{r^2} \f{\p^2 P_{\gamma}}{\p \theta^2}\\ 
	\nn &+& \f{1}{r}\f{\p}{\p r} \l \bar\mu_{\gamma}\, r\, \f{ \p\mr{U_{\gamma}}}{\p r} P_{\gamma}\r + \f{T\Delta\mu_{\gamma}}{r}\f{\p}{\p r} \l  r \cos\theta \f{\p P_{\gamma}}{\p R_x} +  r \sin\theta \f{\p P_{\gamma}}{\p R_y}\r + \f{T\bar\mu_{\gamma}}{r} \f{\p}{\p r} \l r \f{\p P_{\gamma}}{\p r}\r \\
	&+& k_{\gamma\delta} P_{\delta} -  k_{\delta\gamma} P_{\gamma}.
	\eea 
	
	\section{Effective dynamics of the dimer at long timescales }\label{appendix-effectivedynamics}
	
	Integrating \refn{ap:fp_polar1} w.r.t $r$ gives
	\bea 
	\label{ap:fp_moment0}
	\nn \p_t \tilde P_{\gamma} &=& \f{\p}{\p R_x} \l \lla \Delta\mu_{\gamma} \l \f{ \p\mr{U_{\gamma}}}{\p r} - \f{T}{r} \r\rla_{\gamma}\cos\theta - \sin\theta\f{\p}{\p \theta} \lla\f{T\Delta\mu_{\gamma}}{r}\rla_{\gamma} + T \bar\mu_{\gamma} \f{\p \tilde P_{\gamma}}{\p R_x} \r  \\
	\nn &+& \f{\p}{\p R_y} \l \lla \Delta\mu_{\gamma} \l \f{ \p\mr{U_{\gamma}}}{\p r} - \f{T}{r}\r\rla_{\gamma}\sin\theta +  \cos\theta\f{\p}{\p \theta}\lla\f{T\Delta\mu_{\gamma}}{r}\rla_{\gamma}  + T \bar\mu_{\gamma}\f{\p \tilde P_{\gamma}}{\p R_y} \r  \\
	\nn &+& \f{\p }{\p \theta}\l -\sin\theta \f{\p}{\p R_x} + \cos\theta \f{\p}{\p R_y}\r \lla \f{T\Delta\mu_{\gamma}}{r} \rla_{\gamma} + \f{\p^2}{\p \theta^2}\lla \f{T\bar\mu_{\gamma}}{r^2} \rla_{\gamma} \\
	&+& k_{\gamma\delta} \tilde P_{\delta} -  k_{\delta\gamma} \tilde P_{\gamma},
	\eea 
	where $\la \phi \ra_{\gamma} = \int_0^{\infty} dr r \phi\, P_\gamma $ and $\tilde P_\gamma =  \int_0^{\infty} r dr P_\gamma$ corresponds of $\phi = 1$. 
	To obtain the explicit dynamics of $\tilde P_\gamma$ we need to solve for higher moments or $r$, i.e, $\la r^n \ra_\gamma$. To make analytical progress, we take the interaction potential to be harmonic, i.e.,  $\mr{U_{\gamma}} = k_{\gamma}(r-l_{\gamma})^2/2 $. Furthermore, we consider the parameter values for which $(r-l_\gamma)/l_\gamma \ll 1$. With this, expanding $r$ around $l_{\gamma}$, to leading order gives 
	\be 
	\f{T}{r} \approx \f{T}{l_\gamma} \l 1 - \f{r-l_\gamma}{l_\gamma} \r.
	\ee 
	Using this the drift term in \refn{ap:fp_moment0} reads
	\be 
	\f{\p\mr{U_{\gamma}}}{\p r} - \f{T}{r} \approx \tilde k(r - \tilde l_\gamma) ,
	\ee 
	where the renormalized stiffness and length of the spring are $\tilde k_{\gamma} = \l k_{\gamma} + T/l_{\gamma}^2\r$ and $\tilde l_{\gamma} = \l1+T/(k_{\gamma}l_{\gamma}^2+T)\r$.
	Estimates show that $T/k_\gamma l_\gamma^2 << 1$, which gives  $(r-l_\gamma)/l_\gamma \sim \sqrt{T/k_\gamma l_\gamma^2} \ll 1 $. This allows us to further simplify the relations. In the following working to leading order we take $T/r = T/l_{\gamma}$, $\tilde k = k$, and $\tilde l = l$. Substituting this approximation in \refn{ap:fp_moment0} we get
	\bea 
	\label{ap:fp_moment00}
	\nn \p_t \tilde P_{\gamma} &=& \f{\p}{\p R_x} \l \Delta\mu_{\gamma} k_\gamma \l \la r\ra_\gamma - l_\gamma \tilde P_\gamma\r \cos\theta - \sin\theta\f{T\Delta\mu_{\gamma}}{l_{\gamma}}\f{\p \tilde P_\gamma}{\p \theta}  + T \bar\mu_{\gamma} \f{\p \tilde P_{\gamma}}{\p R_x} \r  \\
	\nn &+& \f{\p}{\p R_y} \l \Delta\mu_{\gamma} k_\gamma \l \la r\ra_\gamma - l_\gamma \tilde P_\gamma\r \sin\theta +  \cos\theta\f{T\Delta\mu_{\gamma}}{l_{\gamma}}\f{\p \tilde P_\gamma}{\p \theta}  + T \bar\mu_{\gamma}\f{\p \tilde P_{\gamma}}{\p R_y} \r  \\
	\nn &+& \f{T\Delta\mu_{\gamma}}{l_\gamma}\f{\p }{\p \theta}\l -\sin\theta \f{\p \tilde P_\gamma}{\p R_x} + \cos\theta \f{\p \tilde P_\gamma}{\p R_y}\r  + \f{T\bar\mu_{\gamma}}{l_\gamma^2}\f{\p^2 \tilde P_\gamma}{\p \theta^2}  \\
	&+& k_{\gamma\delta} \tilde P_{\delta} -  k_{\delta\gamma} \tilde P_{\gamma},
	\eea 
	
	The first moment obtained from \refn{ap:fp_polar1}  reads
	\bea 
	\label{ap:fp_moment1}
	\nn \p_t \la r \ra_\gamma &=& \f{\p}{\p R_x} \l \lla \Delta\mu_{\gamma} r \l \f{ \p\mr{U_{\gamma}}}{\p r} - \f{T}{r} \r\rla_\gamma\cos\theta - T\Delta\mu_{\gamma}\sin\theta\f{\p \tilde P_\gamma }{\p \theta} + T \bar\mu_{\gamma}  \f{\p \la r \ra_\gamma}{\p R_x} \r \\
	\nn &+& \f{\p}{\p R_y} \l \lla \Delta\mu_{\gamma}r \l \f{ \p\mr{U_{\gamma}}}{\p r} - \f{T}{r}\r\rla_\gamma\sin\theta +  T\Delta\mu_{\gamma}\cos\theta\f{\p \tilde P_\gamma }{\p \theta}  + T \bar\mu_{\gamma} \f{\p \la r \ra_\gamma}{\p R_y} \r\\
	\nn &+& T\Delta\mu_{\gamma}\f{\p }{\p \theta}\l -\sin\theta \f{\p \tilde P_{\gamma}}{\p R_x} + \cos\theta \f{\p \tilde P_{\gamma}}{\p R_y}\r + \f{\p^2 }{\p \theta^2}  \lla \f{T\bar\mu_{\gamma}}{r} \rla_\gamma \\
	&-&  \bar\mu_{\gamma}\lla \f{ \p\mr{U_{\gamma}}}{\p r} - \f{T}{r}\rla_\gamma - T\Delta\mu_{\gamma} \l   \cos\theta \f{\p \tilde P_{\gamma}}{\p R_x} +  \sin\theta \f{\p \tilde P_{\gamma}}{\p R_y}\r + k_{\gamma\delta} \tilde P_{\delta} -  k_{\delta\gamma} \tilde P_{\gamma},
	\eea 
	Dropping the derivative terms in \refn{ap:fp_moment1}, which lead to higher order derivative terms in \refn{ap:fp_moment0}, and working in the same approximation as \refn{ap:fp_moment00} we get
	\bea 
	\label{ap:moment1}
	\p_t \la r \ra_{\gamma} &=& -   \bar \mu_{\gamma}  k_\gamma \l \la r \ra_\gamma - l_\gamma \tilde P_\gamma \r- T\Delta\mu_{\gamma}\l  \cos\theta \f{\p \tilde P}{\p R_x} + \sin\theta \f{\p \tilde P}{\p R_y}\r + k_{\gamma\delta} \la r \ra_{\delta} -  k_{\delta\gamma} \la r \ra_{\gamma}.
	\eea 
	Since the rate of switching between the bound and unbound states is much slower than the rate of switching between the intermediate bound states and the separation relaxation timescales, the separation dynamics as given by \refn{ap:moment1} simplifies to the following equations for the three states:
	\bea
	\label{ap:sep1}
	\p_t\la r\ra_1 &=& - \bar \mu_1   k_1(\la r \ra_1 - l_1\tilde P_1) - T\Delta\mu_1\, {\bf \hat n} \cdot \nabla_{\bf R} \tilde P_1 ,\\ 
	\label{ap:sep2}
	\p_t \la r\ra_2 &=& - \bar \mu_2   k_2(\la r \ra_2 - l_2 \tilde P_2) - T\Delta\mu_2\, {\bf \hat n} \cdot \nabla_{\bf R} \tilde P_2 + k_{23} \la r\ra_3 -  k_{32} \la r\ra_2, \quad \\
	\label{ap:sep3}
	\p_t \la r\ra_3  &=&-\bar \mu_3  k_3(\la r\ra_3 - l_3 \tilde P_3) - T\Delta\mu_3\, {\bf \hat n} \cdot \nabla_{\bf R} \tilde P_3 + k_{32} \la r\ra_2 - k_{23} \la r\ra_3.
	\eea   
	
	Substituting the steadystate values of separation back in \refn{ap:fp_moment00} the dynamics is state one as
	\bea 
	\label{ap:fp_state1}
	\nn \p_t \tilde P_u &=& \f{\p}{\p R_x} \l  - \f{T\Delta\mu}{l_u}\sin\theta\f{\p  \tilde P_u}{\p \theta} + T\l \bar \mu_u - \f{\Delta\mu^2_u}{\bar \mu_u} \cos^2\theta \r \f{\p \tilde P_u}{\p R_x} - \f{T\Delta\mu^2_u}{\bar \mu_u} \sin\theta\cos\theta\f{\p \tilde P_u}{\p R_y}\r \\
	\nn &+& \f{\p}{\p R_y} \l \f{T\Delta\mu}{l_u}\cos\theta\f{\p\tilde P_u}{\p \theta}  + T \l \bar \mu_u - \f{\Delta\mu^2_u}{\bar \mu_u} \sin^2\theta \r \f{\p \tilde P_u}{\p R_y} - \f{T\Delta\mu^2_u}{\bar \mu_u} \sin\theta\cos\theta\f{\p\tilde P_u}{\p R_x}\r \\
	&+& \f{T\Delta\mu}{l_u}\f{\p }{\p \theta} \l -\sin\theta \f{\p \tilde P_u }{\p R_x} + \cos\theta \f{\p \tilde P_u}{\p R_y}\r  +  \f{T\bar \mu}{l_u^2}\f{\p^2 \tilde P_u}{\p \theta^2} - k_{bu} \tilde P_u + k_{ub} \tilde P_b,
	\eea 
	where $\tilde P_u = \tilde P_1$, $\tilde P_b = \tilde P_2 + \tilde P_3$, $k_{ub} = (k_{23}k_{12}+k_{32}k_{13})/(k_{23}+k_{32})$ and $k_{bu}=(k_{21} + k_{31})$. 
	The dynamics in the bound state obtained after substituting the steady state separation obtained from \refn{ap:sep2} and \refn{ap:sep3} in \refn{ap:fp_moment00} reads 
	\bea 
	\label{ap:fp_stateb}
	\nn\p_t \tilde P_b &=& \f{\p}{\p R_x} \l v_b\cos\theta - \lla\f{T\Delta\mu}{l}\rla_b\sin\theta\f{\p}{\p \theta} + T\l\la \bar \mu\ra_b -  \delta\mu_b\cos^2\theta \r\f{\p}{\p R_x} - T\delta\mu_b \sin\theta\cos\theta\f{\p}{\p R_y}\r \tilde P_b \\
	\nn &+& \f{\p}{\p R_y} \l v_b\sin\theta +  \lla \f{T\Delta\mu}{l}\rla_b\cos\theta\f{\p}{\p \theta}  + T \l \la\bar \mu\ra_b - \delta\mu_b \sin^2\theta \r \f{\p}{\p R_y} - T\delta\mu_b \sin\theta\cos\theta\f{\p}{\p R_x}\r \tilde P_b \\
	&+& \lla \f{T\Delta \mu}{l} \rla_b \f{\p }{\p \theta}\l -\sin\theta \f{\p \tilde P_b}{\p R_x} + \cos\theta \f{\p \tilde P_b}{\p R_y}\r + \lla \f{T\bar \mu}{l^2} \rla_b \f{\p^2 \tilde P_b}{\p \theta^2} + k_{bu} \tilde P_u - k_{ub} \tilde P_b,
	\eea
	where $\la \phi\ra_b = (k_{23}\phi_2 + k_{32}\phi_3)/(k_{32}+k_{23})$. 
	\bea 
	\label{ap:vel}
	v_b &=& \f{2k_{23}k_{32}\tilde k_2 \tilde k_3\mu_a(\bar\mu_2-\bar\mu_3)\delta l}{(k_{23}+k_{32})\l \bar\mu_2 \tilde k_2\bar\mu_3 \tilde k_3 + k_{23}\bar\mu_2 \tilde k_2  + k_{32} \bar \mu_3 \tilde k_3 \r}, 
	\eea 
	and
	\be
	\delta \mu_b = \f{\Delta \mu_3^2 (\bar \mu_2 k_2 + k_{32})k_{32} k_3 + \Delta \mu_2^2 (\bar \mu_3 k_3 + k_{23})k_{23} k_2 + (k_2 + k_3)k_{32}k_{23}\Delta \mu_2 \Delta \mu_3 }{(k_{23}+k_{32})\l \bar\mu_2 k_2\bar\mu_3 k_3 + k_{23}\bar\mu_2 k_2  + k_{32} \bar \mu_3 k_3 \r}.
	\ee 
	For $\bar \mu_2 k_2, \bar \mu_3 k_3 \gg k_{32},k_{23}$ we get 
	\be
	\delta \mu_b = \f{\Delta \mu_3^2 }{\bar \mu_3} d_3 + \f{\Delta \mu_2^2 }{\bar \mu_2} d_2, 
	\ee 
	where $d_2 = k_{23}/(k_{23}+k_{32})$ and $d_3 = k_{32}/(k_{23}+k_{32})$.
	In the opposite limit, $\bar \mu_2 k_2, \bar \mu_3 k_3 \ll k_{32},k_{23}$, to zeroth order  we get 
	\be
	\delta \mu_b = \f{\Delta \mu_3^2  k_{32}^2 k_3 + \Delta \mu_2^2 k_{23}^2 k_2 + (k_2 + k_3)k_{32}k_{23}\Delta \mu_2 \Delta \mu_3 }{(k_{23}+k_{32})\l  k_{23}\bar\mu_2 k_2  + k_{32} \bar \mu_3 k_3 \r}
	\ee 
	\be
	\delta \mu_b = \f{(\Delta\mu_3 k_{32}k_3 + )(\Delta \mu_2 k_{23} + \Delta \mu_3 k_{23})}{(k_{23}+k_{32})\l  k_{23}\bar\mu_2 k_2  + k_{32} \bar \mu_3 k_3 \r}
	\ee

	\section{Effective diffusion constant}\label{appendix-DiffConst}
	
	In each state, the particle has an effective diffusion constant. 
	
	For this Fokker-Planck equation 	
	the moments $p^\gamma_{ij} = \int_{0}^{2\pi} d\theta \cos^i(\theta)\sin^j(\theta) \tilde P_\gamma(\bf {R},\theta)$, obtained from \refn{ap:fp_state1} and \refn{ap:fp_stateb} are given by
	\bea 
	\label{ap:p00}
	\p_t p_{00}^\gamma &=& \f{\p}{\p R_x} \l \l v + \lla \f{T\Delta\mu}{l}\rla_\gamma\r p_{10}^\gamma + T\la \bar \mu \ra  \f{\p p_{00}^\gamma}{\p R_x}- T\lla\delta\mu \rla \f{\p p_{20}^\gamma}{\p R_x}\r  - 2T\lla \delta\mu \rla \f{\p^2p_{11}^\gamma }{\p R_x \p R_y}\\
	\nn &+&\f{\p}{\p R_y} \l \l v + \lla\f{T\Delta\mu}{l}\rla_\gamma\r p_{01}^\gamma + T\la \bar \mu \ra  \f{\p p_{00}^\gamma}{\p R_y}- T\lla \delta\mu \rla \f{\p p_{02}^\gamma}{\p R_y}\r ,
	\eea 
	Thus we see that to obtain the diffusion equation for the dimer we need to solve for the higher moments. For the rotations diffusion timescales of the order of or smaller than the binding-unbinding kinetics we use moment closure approximation. We consider the moments up to which we second terms that contribute terms with a second derivative or less to the zeroth moment. Thus we keep terms only up to first order in derivatives for the moments $p_{10}^\gamma$ and $p_{01}^\gamma$, and only zeroth order terms in $p_{20}^\gamma$, $p_{02}^\gamma$, and $p_{11}^\gamma$. In this approximation we get 
	\bea 
	\label{ap:p10}
	\nn\p_t p_{10}^\gamma &=& \f{\p}{\p R_x} \l \l v - \lla \f{T\Delta\mu}{l}\rla_\gamma\r p_{20}^\gamma + \lla\f{2T \Delta\mu}{l}\rla_\gamma (2p_{20}^\gamma - p_{00}^\gamma) \r +  \f{\p}{\p R_y} \l \l v_b - \lla \f{T\Delta\mu}{l}\rla_\gamma\r p_{11}^\gamma + \lla\f{4T \Delta\mu}{l}\rla_\gamma p_{11}^\gamma \r\\
	&-& \lla\f{T \bar \mu}{r^2} \rla p_{10}^\gamma  +\mr{higher\, order\, derivatives}, \\
	\label{ap:p01}
	\nn\p_t p_{01}^\gamma &=& \f{\p}{\p R_x} \l \l v - \lla \f{T\Delta\mu}{l}\rla_\gamma\r p_{11}^\gamma + \lla\f{4T \Delta\mu}{l}\rla_\gamma p_{11}^\gamma \r +  \f{\p}{\p R_y} \l \l v_b - \lla \f{T\Delta\mu}{l}\rla_\gamma\r p_{20}^\gamma + \lla\f{2T \Delta\mu}{l}\rla_\gamma (2p_{20}^\gamma - p_{00}^\gamma) \r\\
	&-& \lla \f{T \bar \mu}{l^2}\rla_\gamma p_{01}^\gamma  + \mr{higher\, order\, derivatives}.
	\eea 
	For this, we need to keep the zeroth order term in $p_{11}^\gamma$.
	\bea 
	\p_t p_{20}^\gamma  &=& -\lla\f{4T \bar \mu}{l^2}\rla_\gamma (2p_{20}^\gamma-p_{00}^\gamma) + \mr{higher\, order\, derivatives},\\
	\p_t p_{02}^\gamma  &=& -\lla\f{4T \bar \mu}{l^2}\rla_\gamma (2p_{02}^\gamma-p_{00}^\gamma) + \mr{higher\, order\, derivatives},\\
	\p_t p_{11}^\gamma &=& - \lla\f{4T \bar \mu}{l^2}\rla_\gamma p_{11}^\gamma + \mr{higher\, order\, derivatives}.
	\eea 
	At steady state,  we get $p_{11}^\gamma=0$, $p_{20}^\gamma =p_{02}^\gamma = p_{00}^\gamma/2$. Substituting values of $p_{02}^\gamma$ and $p_{20}^\gamma$ in the equation for $p_{01}^\gamma$ and $p_{10}^\gamma$ gives
	\bea 
	\lla\f{T \bar \mu}{l^2}\rla_\gamma p_{10}^\gamma = \f{\p}{\p R_x} \l \f{1}{2}\l v - \lla \f{T\Delta\mu}{l}\rla_\gamma\r p_{00}^\gamma  \r , \\
	\lla\f{T \bar \mu}{l^2}\rla_\gamma p_{01}^\gamma= \f{\p}{\p R_y} \l \f{1}{2}\l v - \lla \f{T\Delta\mu}{l}\rla_\gamma\r p_{00}^\gamma \r.
	\eea 
	
	\bea 
	\p_t p_{00}^\gamma &=& \l \f{v^2}{2\la T\bar\mu/l^2\ra_\gamma} + T\l \la \bar \mu \ra_\gamma  - \f{1}{2\la \bar\mu/l^2\ra_\gamma}\lla \f{\Delta\mu}{l}\rla^2_\gamma  - \f{1}{2}\delta\mu_\gamma \r \r \nabla^2 p_{00}^\gamma.
	\eea 
	In the unbound state  we get
	\bea 
	\p_t p_{00}^u &=& T\l \f{4\mu_a \mu_1}{\mu_a + \mu_1}  \r \nabla^2 p_{00}^u.
	\eea 
	In the bound state, we get
	\bea 
	\p_t p_{00}^b &=& \l \f{v_b^2}{2\la T\bar\mu/l^2\ra_b} + T\l \la \bar \mu_b \ra  - \f{1}{2\la \bar\mu/l^2\ra_b}\lla \f{\Delta\mu}{l}\rla^2_b  - \f{1}{2}\delta\mu_b \r \r \nabla^2 p_{00}^b.
	\eea

\section{Deterministic dynamics with hydrodynamic interaction}\label{appendix-DetDyn}

The dynamics of particles including hydrodynamic interaction reads:
\be
	\label{eq:langevin_HI}
	\f{d {\bf R}_\mr{A}}{dt} = (\mu_\mr{A} - \mu_{AB}) \nabla_r \mr{U_{\gamma}}(r) \quad \mr{and} \quad
	\f{d {\bf R}_\mr{B}}{dt} = -(\mu_{\gamma} - \mu_{AB}) \nabla_r \mr{U_{\gamma}}(r) ,
\ee
where $\mu_{AB}$ is the hydrodynamic interaction tension between the two particles. The dynamics of mean position and separation variable reads. 
\be
\f{d {\bf R}}{dt} = (\mu_A - \mu_\gamma) k_\gamma (r-l_\gamma) \quad \mr{and} \quad \f{dr}{dt} = -(\mu_A + \mu_\gamma - 2\mu_{AB} )k_\gamma (r-l_\gamma).
\ee 
At time $t=0$ the potential changes from  $\mr{U}_3$ to $\mr{U}_2$, and the mobility changes from $\mu_3$ to $\mu_2$. In the limit $\mu_a k_{\gamma},\mu_{\gamma} k_{\gamma}\gg \tau$, the dimer relaxes to the new length $l_2$. 
	From \refn{eq:pos_dimer} of the main text we get that the dimer  moves by distance $\Delta R_3 = (l_2-l_3)(\mu_A -\mu_2)/(\mu_A + \mu_2 - 2\mu_{AB})$ .
	At the time $t = \tau$ the potential and the mobility switch back to  $\mr{U}_3$ and $\mu_3$, respectively. The dimer now relaxes to the new equilibrium length $l_3$, and the dimer moves a distance $\Delta R_2 = (l_3-l_2)(\mu_A - \mu_3)/(\mu_A + \mu_3 - 2\mu_{AB})$. The total displacement in one cycle of time $2\tau$ is $\Delta R = \Delta R_3 + \Delta R_2$,
 \be 
 \Delta R = (l_2-l_3)\l \f{(\mu_A -\mu_2)}{(\mu_A + \mu_2 - 2\mu_{AB})} - \f{(\mu_A - \mu_3)}{(\mu_A + \mu_3 - 2\mu_{AB})}\r.
 \ee 
 The velocity $v = \Delta R/2\tau$ reads
   \be 
 v = \f{(l_2-l_3)}{\tau} \f{ (\mu_A + \mu_{AB})(\mu_2 -\mu_3)}{(\mu_A + \mu_2 - 2\mu_{AB})(\mu_A + \mu_3 - 2\mu_{AB})} .
 \ee 
 This is qualitatively similar to the expression \refn{eq:det_v} obtained without the hydrodynamic interaction. Since $\mu_{AD} > 0$, the inclusion of hydrodynamic interactions leads to an increase in the velocity. 
	
\end{document}